\documentclass{sistedes}
\usepackage[T1]{fontenc}
\usepackage{graphicx}
\usepackage{multirow} 
\usepackage{pdflscape} 
\usepackage{hyperref} 
\usepackage{xcolor} 
\usepackage{subcaption} 
\usepackage{amssymb} 
\usepackage{pifont}
\usepackage{tcolorbox}
\usepackage{xspace}
\usepackage[colorinlistoftodos,textsize=small]{todonotes} 
\usepackage{tabularx} 

\newcolumntype{C}{>{\centering\arraybackslash}X}
\newcolumntype{L}{>{\raggedright\arraybackslash}X}

\newcommand{\solutionName}{Pricing4SaaS}

\begin{document}
\title{Pricing4SaaS: a suite of software libraries for pricing-driven feature toggling
}
%
%
\author{Alejandro García-Fernández\orcidID{0009-0000-0353-8891} \and
José Antonio Parejo\orcidID{0000-0002-4708-4606} \and
Pablo Trinidad \orcidID{0000-0002-1320-2424}\and
Antonio Ruiz-Cortés\orcidID{0000-0001-9827-1834}}

\titlerunning{\solutionName: a suite of software libraries for pricing-driven feature toggling}

\authorrunning{García-Fernández et al.}
%
\institute{SCORE Lab, I3US Institute, Universidad de Sevilla, Espa\~na \\
\email{\{agarcia29,japarejo,ptrinidad,aruiz\}@us.es}}
\maketitle              
\begin{abstract}

As the digital marketplace evolves, the ability to dynamically adjust or disable features and services in response to market demands and pricing strategies becomes increasingly crucial for maintaining competitive advantage and enhancing user engagement. This paper introduces a novel suite of software libraries named \solutionName, designed to facilitate the implementation of pricing-driven feature toggles in both the front-end and back-end of SaaS systems, and discuss its architectural design principles. Including Pricing4React for front-end and Pricing4Java for back-end, the suite enables developers a streamlined and efficient approach to integrating feature toggles that can be controlled based on pricing plans, emphasizing centralized toggle management, and secure synchronization of the toggling state between the client and server. We also present a case study based on the popular Spring PetClinic project to illustrate how the suite can be leveraged to optimize developer productivity, avoiding technical debt, and improving operational efficiency.

\end{abstract}

\section{Introduction}

In the dynamic landscape of Software as a Service (SaaS) development, feature toggling has emerged as a powerful strategy for managing and deploying new functionalities in a controlled and incremental manner \cite{FOWLER2023}. This approach allows developers to enable or disable features at runtime without deploying new code, providing a flexible mechanism for A/B testing, canary releases, and minimizing downtime. The ability of subscription based licensing with multiple pricing plans enables a predictable revenue stream for SaaS providers, while offering flexibility and scalability to users \cite{Jiang2009}. In this context, feature toggling is an essential mechanism to adapt the interface and functionality of SaaS to the specific users' pricing plan, given that the adoption of subscription models has become the most popular software licensing mechanism in SaaS systems. 

However, implementing feature toggling becomes increasingly complex and challenging when the variables and conditions that govern the toggles are subject to frequent changes. This complexity is further amplified in environments where features are directly influenced by rapidly evolving market conditions, user preferences, or pricing strategies. The crux of the problem lies in the need for a robust system that can not only manage a multitude of feature toggles but also adapt to changes in controlling variables with minimal latency and overhead in distributed environments. Traditional feature toggle systems with manual implementation of variables evaluation and toggling conditions, and hard-coded toggle configuration, often fall short in scenarios where toggles must respond dynamically to external factors. This limitation can lead to significant technical debt, decreased system performance, and a higher risk of errors, ultimately affecting the user experience and organizations' ability to respond to market demands.

This paper introduces a suite of libraries designed specifically to address these challenges by facilitating pricing-driven feature toggling. Our approach leverages a centralised pricing configuration to connect every pricing-dependent toggle point within the code base, allowing SaaS to rapidly adapt to changes in pricing strategies and market conditions autonomously. By providing a demonstration of our libraries' capabilities, we aim to showcase how our solution not only mitigates the complexities associated with frequent changes in feature toggling conditions but also enhances operational efficiency and market responsiveness for teams creating SaaS.

\vspace{-0.4cm}

\section{Pricing4SaaS Architecture and Features}

\vspace{-0.2cm}Pricing4SaaS is a set of libraries that ease the integration of pricing plans within SaaS systems, elevating it to a first-class citizen. It relies on the feature toggles technology, which allows to modify a system's behavior without changing the code \cite{FOWLER2023}. In Fowler's terminology, Pricing4SaaS apply ``Inversion of Decision'' to make the toggling router configure each toggling variable for us in the corresponding toggling point, and add an additional layer of indirection so that developers can customize the computation of the relevant state variables for its domain and provide them to the toggling router.

\begin{figure}[htb]
    \centering
    \makebox[\textwidth][c]{\includegraphics[width=1\textwidth,trim={0 2.2cm 0 0.1cm},clip]{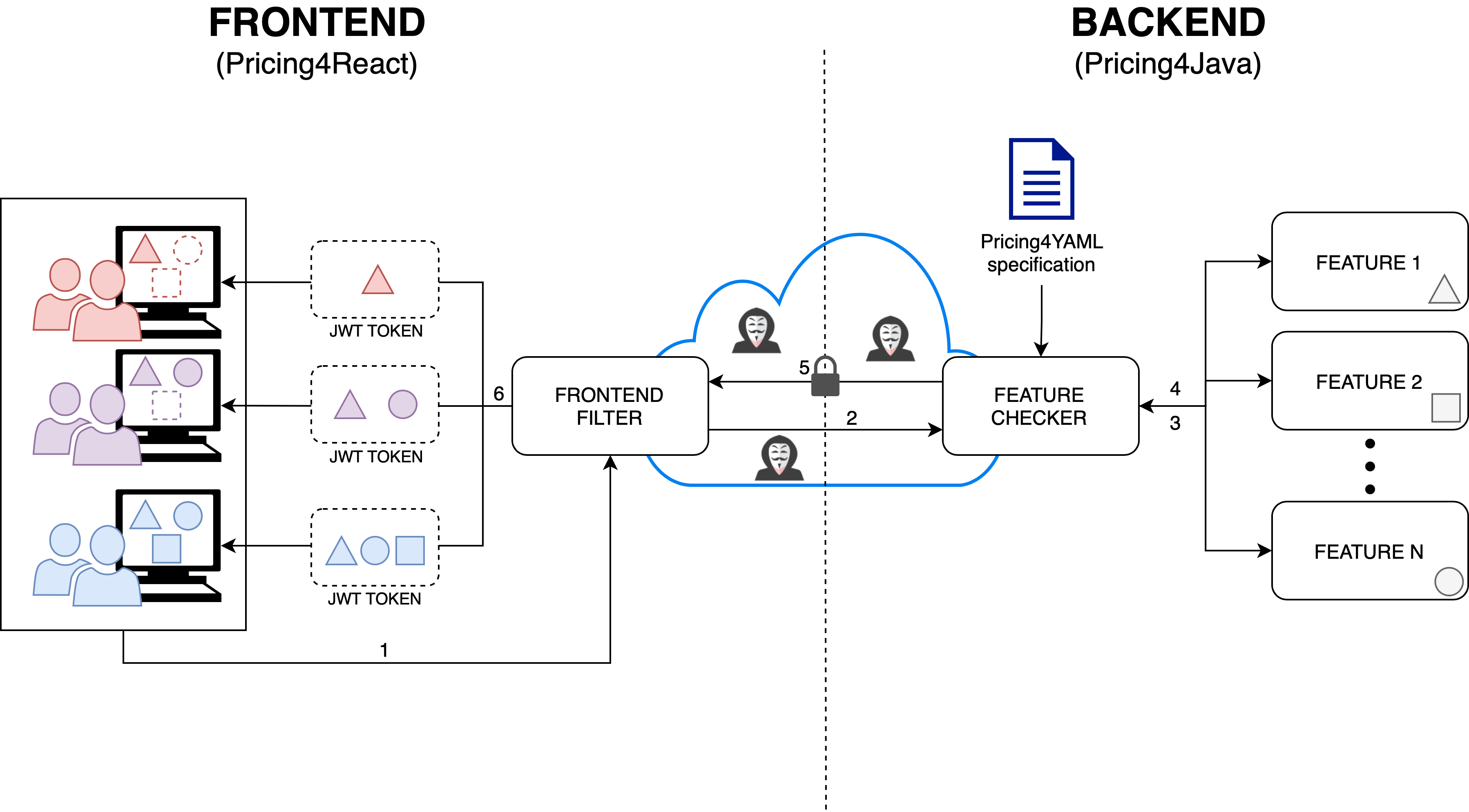}}
    \caption{Architectural design of \solutionName}
    \label{fig:pricingplansArchitecture}
    \vspace{-0.4cm}
\end{figure}

Figure \ref{fig:pricingplansArchitecture} represents the process of a request within the {\solutionName} architecture. The journey begins at the front-end (1), where users interact with the system's User Interface (UI). Each user's session is associated with a JWT token that encodes permissions and subscription details. When a client uses a feature $F$, the front-end sends a request to the back-end, carrying his JWT token (2). In the back-end, \textit{Pricing4Java} acts as the \textbf{Feature Checker} middleware component (3) and intercepts the request to evaluate the client's JWT token against a \textbf{Yaml4SaaS} specification, a YAML-based syntax that models the rules and conditions of a pricing and determine the features available under certain circumstances \cite{GARCIA23}. If $F$ is enabled for the user, and the token signature is correct, which means it has not suffer any alterations during the transmission over the network, the back-end processes the request and returns the required data again to the Feature Checker (4), which wraps an updated version of the feature evaluation into the JWT if it has changed, e.g. limit of $F$ has been reached by the user after using it. Finally, the response is sent to the front-end (5), where \textit{Pricing4React} process the new JWT, if exists, and updates the UI according with the feature evaluations received (6), ensuring a seamless user experience tailored to user's subscription. In this case, the button activating $F$ should disappear.

\paragraph{Pricing4React}

The library\footnote{Pricing4React GitHub repository can be found \href{https://github.com/isa-group/pricingplans-react}{here}} serves as a main component in the front-end ecosystem, enabling feature toggling within a React-based user interface. It provides a set of components that allow to use the real-time feature evaluation received from the back-end through a JWT to display, hide, or alter UI elements, ensuring that the user experience is congruent with their subscription level. 

This setup is provided by a filter component that leverages JWT to receive from the back-end a secure and authenticated payload, which includes the evaluation of various features pertinent to the requesting user. This JWT, stored within the browser's local storage, equips the front-end with essential information about feature availability and usage limitations, as defined by the user's current plan. Despite the JWT management can be implemented without using the filter included in the library, we recommended its use, as it allows for seamless on-the-fly pricing updates to the UI without requiring manual intervention, streamlining the user experience in a dynamic pricing-driven environment.

\begin{figure}
\vspace{-0.5cm}
    \includegraphics[width=\textwidth]{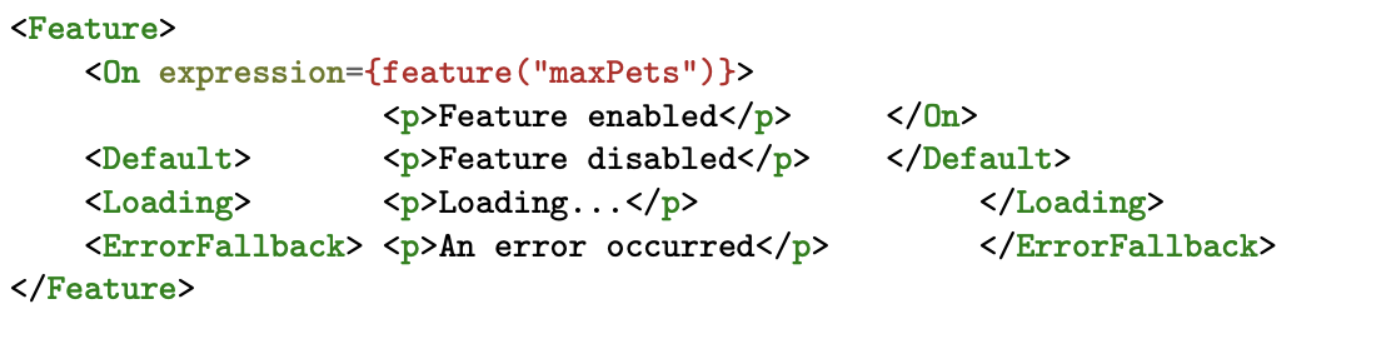}
    \vspace{-0.8cm}
    \caption{Declaration of a dynamically enabled feature using \textit{Pricing4React}}
    \label{fig:featureTogglingReact}
    \vspace{-0.4cm}
\end{figure}

The cornerstone of the remain logic is the \textbf{Feature} component, which can contain up to four subcategories: On, Default, Loading, ErrorFallback; that determine which component is rendered based on the evaluation result (Figure \ref{fig:featureTogglingReact}).
\vspace{-0.3cm}
\vspace{-0.4cm}
\paragraph{Pricing4Java}

This Java-based toolkit\footnote{Pricing4Java GitHub repository can be found \href{https://github.com/isa-group/pricingplans-4j}{here}} is designed to enhance the server-side functionality by enabling the seamless integration of pricing plans into the application logic. The package provides a suite of components that are predicated on the Pricing4YAML syntax \cite{GARCIA23}. This specification facilitates the definition of system pricing and its features alongside their respective evaluation expressions, grouping them within plans and add-ons, as well as establishing usage limits.

The cornerstone of the package is the \textbf{PricingContext}, an abstract class that empowers developers to tailor the library's capabilities to the pricing framework of their applications. Once it is configured, it unlocks a suite of business logic tools, such as: a service for managing the pricing plan or custom annotations that automate the back-end validation of the pricing rules.

In addition to these components, the package can also manage the JWTs that contain the result of the feature evaluation for an user alongside the authentication data defined within the application. Thus, we can automatically keep the feature evaluation data up to date without needing to implement a polling mechanism that overloads the server with additional API calls.



\section{Demonstration and Use Cases}




In order to assure the validity of our approach, we have successfully applied it into a \href{https://github.com/isa-group/petclinic-react}{Spring+React version of the PetClinic application}\footnote{A demo can be found \href{https://youtu.be/gS_577KjrXg}{here}.}, a sample veterinary clinic management system designed to illustrate the functionality and features of a particular software framework or technology. 

\begin{figure}
    \centering
    \includegraphics[width=\textwidth]{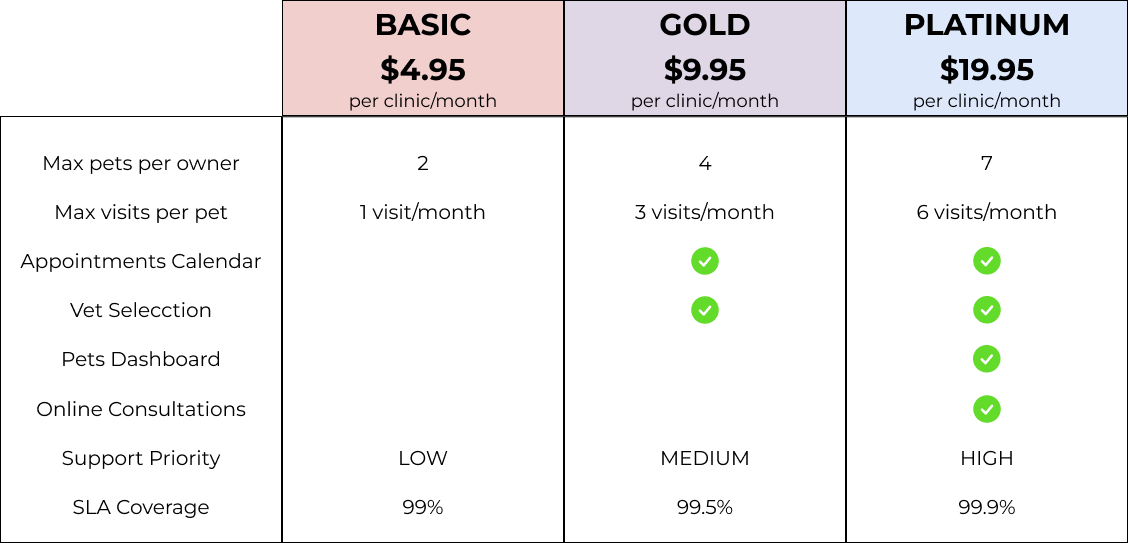}
    \caption{PetClinic's pricing}
    \label{fig:pricingPetclinic}
\end{figure}

Central to this application is a dynamic pricing model that controls access to various features (Figure \ref{fig:pricingPetclinic}). The system delineates four user roles: administrators, clinic owners, veterinarians, and pet owners, with clinic owners empowered to subscribe multiple clinics under different pricing plans simultaneously. Notably, it is the pet owners who experience the direct impact of these pricing constraints.

The integration of Pricing4SaaS into PetClinic achieves several key objectives: i) it facilitates the modeling of the application's pricing structure, enabling the delineation and management of access levels to the application's features; ii) offers administrators a suite of tools directly through the user interface, allowing for the seamless addition of new pricing plans or the adjustment of existing ones. Such changes are applied on real-time within the application, without the need of a developer to manually apply them. iii) Pricing4SaaS automates the management of pricing-driven feature toggles, ensuring that feature access across the platform dynamically responds to changes in subscription levels. Lastly, iv) it streamlines the back-end validation process for pricing limits, enhancing the overall efficiency and reliability of the system.

This implementation does not only prove the practicality and versatility of the Pricing4SaaS suite but also highlights their potential to significantly improve the management and operation of SaaS platforms. By enabling dynamic, pricing-driven feature toggling and simplifying administrative tasks, these libraries present a valuable resource for developers and administrators aiming to tailor their services more closely to user needs and market demands.

\section{Future work}
\label{sec:futurework}
\vspace{-0.25cm}
Several challenges remain for future work, but the main target is to design a UI component in Pricing4React that establish an actual common language between business decision-makers and developers, allowing non-technical users to apply pricing-driven changes, e.g adding a new plan, into the system without actually modifying the source code. We would like this update to also include some new design components, such as: a default pricing view, user subscription manager, etc. In the long term, we plan to develop equivalent libraries for other technologies, such as angular or svelte in front-end, or python/flask in back-end.

\vspace{-0.25cm}
\paragraph{Acknowledgements} 
Authors are thankful to Pedro Gonzalez Marcos for his support in recording the demo attached to this work. This work has been partially supported by grants 
PID2021-126227NB-C21, and 
PID2021-126227NB-C22      
funded by MCIN/AEI/ 10.13039/501100011033/FEDER and European Union ``ERDF a way of making Europe'';
and %
TED2021-131023B-C21 and 
TED2021-131023B-C22     
funded by MCIN/AEI/10.13039/501100011033 and European Union ``NextGenerationEU''/
PRTR;

%
%
%

\vspace{-0.3cm}
 \bibliographystyle{splncs04}
 \bibliography{references}

\begin{thebibliography}{1}
\providecommand{\url}[1]{\texttt{#1}}
\providecommand{\urlprefix}{URL }
\providecommand{\doi}[1]{https://doi.org/#1}

\bibitem{FOWLER2023}
Fowler, M.: Feature toggles (aka feature flags), \url{https://martinfowler.com/articles/feature-toggles.html}

\bibitem{GARCIA23}
Garc{\'i}a-Fern{\'a}ndez, A., Parejo, J.A., Ruiz-Cort{\'e}s, A.: {Pricing4SaaS} - supplementary material (2023). \doi{10.5281/zenodo.10292553}

\bibitem{Jiang2009}
Jiang, Z., Sun, W., Tang, K., Snowdon, J., Zhang, X.: A pattern-based design approach for subscription management of software as a service. 2009 Congress on Services - I pp. 678--685 (2009)

\end{thebibliography}

\end{document}